\documentstyle[pra,aps,floats]{revtex}

\input{epsf}

\newcommand{\bfone}{{\bf 1}}
\newcommand{\bfC}{{\bf C}}

\newcommand{\bfp}{{\bf p}}
\newcommand{\bfP}{{\bf P}}

\newcommand{\bfu}{{\bf u}}
\newcommand{\bfv}{{\bf v}}
\newcommand{\bfw}{{\bf w}}
\newcommand{\bfx}{{\bf x}}
\newcommand{\bfz}{{\bf z}}

\newcommand{\bfgamma}{{\mbox{\boldmath $\gamma$}}}
\newcommand{\bflambda}{{\mbox{\boldmath $\lambda$}}}
\newcommand{\bfnabla}{{\mbox{\boldmath $\nabla$}}}
\newcommand{\calD}{{\hat{\cal D}}}
\newcommand{\calE}{{\cal E}}
\newcommand{\eps}{{\varepsilon}}
\newcommand{\feq}{f^{(\mbox{\scriptsize eq})}}
\newcommand{\fnt}{f^{(\mbox{\scriptsize $0$})}}
\newcommand{\Fnn}[1]{F^{(\mbox{\scriptsize $#1$})}}
\newcommand{\Fnt}{\Fnn{0}}
\newcommand{\sgn}{{\mbox{sgn}}}

\begin{document}

\title{
  \begin{flushleft}
    {\footnotesize To appear, {\it Bras. J. Phys.} (1999)}\\
    {\footnotesize BU-CCS-981201}\\[0.5cm]
  \end{flushleft}
  \bf Navier-Stokes Equations for Generalized Thermostatistics}
\author{
  Bruce M. Boghosian\\
  {\footnotesize Center for Computational Science, Boston University,
    3 Cummington Street, Boston, MA 02215, U.S.A.}\\
  {\footnotesize{\tt bruceb@bu.edu}}\\[0.3cm]
  }
\date{December 7, 1998}
\maketitle

\begin{abstract}
Tsallis has proposed a generalization of Boltzmann-Gibbs
thermostatistics by introducing a family of generalized nonextensive
entropy functionals with a single parameter $q$.  These reduce to the
extensive Boltzmann-Gibbs form for $q=1$, but a remarkable number of
statistical and thermodynamic properties have been shown to be
$q$-invariant -- that is, valid for any $q$.  In this paper, we address
the question of whether or not the value of $q$ for a given viscous,
incompressible fluid can be ascertained solely by measurement of the
fluid's hydrodynamic properties.  We find that the hydrodynamic
equations expressing conservation of mass and momentum are
$q$-invariant, but that for conservation of energy is not.  Moreover, we
find that ratios of transport coefficients may also be $q$-dependent.
These dependences may therefore be exploited to measure $q$
experimentally.
\end{abstract}

\vspace{0.2truein}

\par\noindent {\bf Keywords}: Generalized thermostatistics, Tsallis
entropy, Chapman-Enskog expansion, Navier-Stokes equations.

\section{Introduction}
\subsection{Motivation and Historical Background}

The concept of {\it extensivity} is introduced early in most textbooks
on thermodynamics and statistical physics.  The requirement that the
entropy be additive establishes the form of the Boltzmann-Gibbs
distribution via a straightforward argument.  Recently,
Tsallis~\cite{bib:tsallis} has proposed a generalization of
Boltzmann-Gibbs thermostatistics by introducing a family of generalized
entropy functionals with a single parameter $q$.  The proposed
generalization is best described by the following two axioms:
\newtheorem{axiom}{Axiom}
\begin{axiom}
The entropy functional associated with a probability distribution
$f(\bfz)$ is
\begin{equation}
S_q[f]\equiv\frac{k_B}{q-1}
\int d\bfz\;
\left\{
f(\bfz) - \left[ f(\bfz)\right]^q
\right\}.
\label{eq:ax1}
\end{equation}
\end{axiom}
\begin{axiom} The experimentally measured value of a phase function
$g(\bfz)$ is given by the $q$-expectation value,
\begin{equation}
G_q[f]\equiv\int d\bfz\;
\left[ f(\bfz)\right]^q g(\bfz).
\label{eq:ax2}
\end{equation}
\end{axiom}
From the first axiom, we note that $S_q[f]$ reduces to the
Boltzmann-Gibbs entropy in the limit as $q\rightarrow 1$,
\begin{equation}
S_1[f] = \lim_{q\rightarrow 1}\frac{k_B}{q-1}
\int d\bfz\;
\left\{
f(\bfz) - \left[ f(\bfz)\right]^q
\right\} =
-k_B\int d\bfz\;
f(\bfz)\ln f(\bfz),
\end{equation}
so the generalized thermostatistics includes the usual one as a special
case.  It differs most notably in that fact that neither the entropy
$S_q[f]$ itself nor the observables $G_q[f]$ are extensive thermodynamic
variables when $q\neq 1$.  In spite of this difference, a remarkable
number of statistical and thermodynamic properties have been shown to be
$q$-invariant -- that is, valid for any $q$ whatsoever.  These include
the convexity of the entropy, the equiprobability of the microcanonical
ensemble, the Legendre-transform structure of
thermodynamics~\cite{bib:leg}, and Onsager Reciprocity~\cite{bib:or}, to
name but a few.  Other familiar properties, such as the
Fluctuation-Dissipation Theorem, are not $q$-invariant, but have simple
and straightforward generalizations to arbitrary $q$~\cite{bib:fd}.  The
implication is that the assumption of extensivity plays a role analogous
to that of the parallel postulate of Euclidean geometry; one can deny it
and still get perfectly self-consistent formulations of thermodynamics
and statistical physics.

Of course, all this would be but an idle (though undeniably interesting)
mathematical exercise unless there were actual physical systems whose
thermostatistics are best described by the generalized form with $q\neq
1$.  The exciting realization in recent years is that there do seem to
be some of these.  A very abbreviated list of examples is:
\begin{itemize}
\item Stellar polytropes (e.g., globular clusters) were long known to
possess kinetic equilibria for which there was no corresponding
hydrodynamic variational principle until Plastino and
Plastino~\cite{bib:pp} showed that such a variational principle was
possible only for $q < 7/9$.
\item Experimental studies of pure-electron plasmas in Penning traps
have indicated that such plasmas turbulently relax to a radial density
profile that does not maximize the Boltzmann-Gibbs entropy.  It has
recently been shown~\cite{bib:bog} that the observed profiles are
consistent with $q=1/2$, or perhaps slightly higher~\cite{bib:ant}.
\item The ubiquity of Levy flights in physics can be
explained~\cite{bib:levya,bib:levyb,bib:levyc} by the fact that they are
universal cumulative distributions -- in the same sense that the Central
Limit Theorem establishes the Gaussian as universal -- arising from the
generalized thermostatistics with $q>5/3$.
\item Experimental observations of the velocity distribution of
electrons undergoing inverse bremsstrahlung absorption give results
consistent with $q\neq 1$~\cite{bib:ions}.
\end{itemize}

Two natural questions arise at this point: What characteristics do
physical systems with $q\neq 1$ have in common?  Is there a way to
predict the value of $q$ for a given physical system?  Currently, there
is more intuition about the first of these questions than the second.
Systems that violate extensivity tend to do so because they have a
long-range interaction potential, long-time memory effects, or a fractal
space-time structure.  The first of these qualities can make surface
effects important, even in the thermodynamic limit.  For example, it is
straightforward to show that the total energy of a system of particles
in $D$ dimensions with interaction potential proportional to
$r^{-\zeta}$ diverges if $\zeta < D$~\cite{bib:gold}.  The second
quality can invalidate the Markovian assumption on which much of our
physical intuition is based.  The third quality can introduce scaling
behavior with dimensionality not equal to that of the embedding space,
and it can invalidate the Ergodic Hypothesis which also figures
prominently in the justification of the Boltzmann-Gibbs distribution.

\subsection{Plan of this Paper}

While it is certainly comforting to find familiar properties of
thermodynamics and statistical mechanics that are $q$-invariant --
because this reinforces our existing intuition -- it is no less
important to clearly identify those features that are demonstrably {\it
not} $q$-invariant.  To abuse our above analogy with noneuclidean
geometry, these are the phenomena that are the analog of the
``triangular excess'' of a polygon, or of the curvature tensor.  The
reason these features are interesting is that they are what allows us to
experimentally distinguish systems with different values of $q$.  At the
level of kinetic theory, this is not difficult: The canonical ensemble
distribution function is $q$-dependent, so its direct
measurement~\cite{bib:ions} could provide a way to experimentally
ascertain $q$.  A more subtle question is whether or not different
values of $q$ can give rise to different {\it hydrodynamic} behavior.
This is the central question that we address in this paper.

We begin by developing and studying in some detail as simple a kinetic
theory as we can imagine: We consider an ideal gas for general values of
$q$.  By ``ideal'' here, we mean only that the particles do not interact
except in point collisions; we make no assumptions about the nature of
those collisions.  An objection that might be raised immediately is that
an ideal gas is not the sort of system for which we would expect a
violation of Boltzmann-Gibbs statistics, since it has no long-range
potential.  As mentioned above, however, there is currently no a priori
way to determine $q$ for a given physical system, so we are free to
mandate that $q\neq 1$ -- at least as a mathematical exercise.  We may
suppose that there is some extenuating circumstance that somehow forces
this ideal gas to have $q\neq 1$.  For example, the particles might well
carry a long-term memory of previous collisions, or their geometrical
arrangement and/or the shape of their container might conspire to lead
to gross violations of the Ergodic Hypothesis.  In any case, there is
some precedent for using this system to illustrate the application of
the generalized thermostatistics: Plastino, Plastino and
Tsallis~\cite{bib:ppt} considered the partition function and equilibrium
properties of this very system, including the $q$-dependence of its
specific heat.  In this paper, we concentrate on the {\it hydrodynamic}
behavior of this system.

We next construct a kinetic theory for our general-$q$ ideal gas.  For
the sake of simplicity, our kinetic theory is based on the
Bhatnager-Gross-Krook (BGK) collision operator, which we generalize to
arbitrary $q$, and for which we derive a $q$-invariant $H$-Theorem.
Here it may be argued that the BGK collision operator is too naive to be
used for our purposes, and we ought to have adopted a treatment based on
the full Boltzmann equation.  In defense of the BGK operator, however,
we note that it is well known to produce the correct {\it form} of the
viscous, compressible Navier-Stokes equations when $q=1$, though the
transport coefficients are different from those derived by the full
Boltzmann treatment.  The {\it ratios} of the transport coefficients,
however, are more robust in this regard; for example, both the BGK and
Boltzmann treatments for $q=1$ yield a ratio of bulk to shear viscosity
of $-2/D$, even though the absolute values of those viscosities are
different.  For these reasons, we stick to the BGK operator in this
paper, focus our attention on robust results such as the {\it form} of
the hydrodynamic equations and the {\it ratios} of the transport
coefficients, and leave the more complicated Boltzmann analysis to
future studies.

We then derive the viscous, compressible hydrodynamic equations obeyed
by the system, using a generalization of Chapman and Enskog's asymptotic
expansion in Knudsen number (the ratio of mean-free path to scale
length)~\cite{bib:ce}.  These equations are the generalization to
arbitrary $q$ of the usual Navier-Stokes equations of hydrodynamics.  We
find that the Navier-Stokes equations expressing conservation of mass
and momentum are $q$-invariant, but that for conservation of energy is
not.  Moreover, we find that ratios of transport coefficients may also
be $q$-dependent.  These dependences may therefore be exploited to
measure $q$ by experiments at the level of hydrodynamics.  Finally, in
the process of our analysis, we show that $q$ has a hard upper bound of
$1+2/(D+2)$ for systems of this sort.

\section{Generalized Hydrodynamic Equilibria}
\subsection{Generalized Thermostatistics}

We first review the construction of the canonical ensemble distribution
function using the generalized thermostatistics~\cite{bib:tsallis}.  We
maximize $S_q[f]$, given by Eq.~(\ref{eq:ax1}), subject to the
preservation of various linear global functionals of $f(\bfz)$.  By
Tsallis' second postulate, Eq.~(\ref{eq:ax2}), these are given by
\begin{equation}
C^i_q[f]\equiv
\int d\bfz\; \left[ f(\bfz)\right]^q\gamma^i(\bfz),
\label{eq:constraints}
\end{equation}
where the index $i$ ranges from $1$ to the number of conserved
quantities $n$.  We are thus led to the variational principle,
\begin{equation}
0 = \delta\left\{S_q[f] - \sum_{i=1}^n\lambda_i C^i_q[f]\right\},
\end{equation}
where the $\lambda_i$'s are Lagrange multipliers.  It is an elementary
exercise to verify that this yields the equilibrium distribution
function
\begin{equation}
\feq (\bfz) =
\left\{
q
\left[
1 + (q-1)\sum_{i=1}^n\frac{\lambda_i}{k_B}\gamma^i(\bfz)
\right]
\right\}^{-\frac{1}{q-1}}.
\end{equation}
The $n$ constants $\lambda_i$ are then determined by the $n$
Eqs.~(\ref{eq:constraints}) which may be written
\begin{equation}
C^i_q =
\int d\bfz\;
\left\{
q
\left[
1+(q-1)\sum_{i=1}^n\frac{\lambda_i}{k_B}\gamma^i(\bfz)
\right]
\right\}^{-\frac{q}{q-1}}\gamma^i(\bfz).
\end{equation}

In passing, we note that a very recently proposed modification to
Tsallis' second axiom~\cite{bib:cant} would {\it normalize} the
$q$-expectation values as follows:
\begin{equation}
G_q^\prime[f]
\equiv
\frac{\int d\bfz\;
\left[ f(\bfz)\right]^q g(\bfz)}
{\int d\bfz\;
\left[ f(\bfz)\right]^q}.
\label{eq:ax2a}
\end{equation}
This formulation has the virtue of making the $q$-expectation value of a
constant equal to that constant.  It has been found by
Abe~\cite{bib:abe} to resolve problems with the finiteness of certain
physical observables for the general-$q$ ideal gas.  It has also been
found by Anteneodo~\cite{bib:antp} to yield the same differential
equation, albeit with renormalized coefficients, for the radial profile
of the pure-electron plasma that was found in earlier
studies~\cite{bib:bog,bib:ant}.  In this work, however, we shall adhere
to the original version of the second axiom.  Most of the convergence
problems encountered in the derivation of the equilibrium properties of
the general-$q$ ideal gas~\cite{bib:ppt} do not appear in our derivation
of the hydrodynamic properties of that system.  So, although the use of
normalized $q$-expectation values would be an interesting modification
to the current study, we leave it for future work.

\subsection{Global Hydrodynamic Equilibria}

We next consider phase space coordinates $\bfz=(\bfx,\bfv)$, where
$\bfx$ denotes position and $\bfv$ denotes velocity.  Because our
particles undergo only point collisions, we consider equilibria that
conserve mass, momentum and kinetic energy.  (We do not have to worry
about the potential energy.)  That is, we fix the $q$-expectation values
of the $n=3$ quantities
\begin{eqnarray}
\gamma^1(\bfz) &=& m\\
\bfgamma^2(\bfz) &=& m\bfv\\
\gamma^3(\bfz) &=& mv^2/2,
\end{eqnarray}
namely
\begin{equation}
\left(
\begin{array}{c}
M_q\\
\bfP_q\\
\calE_q
\end{array}
\right)
\equiv
\left(
\begin{array}{c}
C^1_q\\
\bfC^2_q\\
C^3_q
\end{array}
\right)
\equiv\int d\bfz\;
\left[ f(\bfz)\right]^q
\left(
\begin{array}{c}
\gamma^1(\bfz)\\
\bfgamma^2(\bfz)\\
\gamma^3(\bfz)
\end{array}
\right)
=\int d\bfz\;
\left[ f(\bfz)\right]^q
\left(
\begin{array}{c}
m\\
m\bfv\\
mv^2/2
\end{array}
\right).
\end{equation}
The equilibrium distribution function is then
\begin{equation}
\feq (\bfz) =
\left\{
q
\left[
1+(q-1)\frac{m}{k_B}
\left(\lambda_1+\bflambda_2\cdot\bfv+\lambda_3\frac{v^2}{2}\right)
\right]
\right\}^{-\frac{1}{q-1}}.
\end{equation}
This may be written in the more concise form
\begin{equation}
\feq (\bfz) =
\Lambda\left[
1+(q-1)\frac{m}{2k_BT}
\left|\bfv-\bfu\right|^2
\right]^{-\frac{1}{q-1}},
\label{eq:equil}
\end{equation}
where the three quantities
\begin{equation}
\Lambda
\equiv
\left\{
q
\left[1+(q-1)\frac{m}{k_B}
\left(\lambda_1-\frac{\lambda_2^2}{2\lambda_3}\right)
\right]
\right\}^{-\frac{1}{q-1}}
\end{equation}
\begin{equation}
\bfu\equiv -\frac{\bflambda_2}{\lambda_3}
\end{equation}
and
\begin{equation}
T
\equiv
\frac{1}{\lambda_3}
\left[1+(q-1)\frac{m}{k_B}
\left(\lambda_1-\frac{\lambda_2^2}{2\lambda_3}\right)
\right]
\label{eq:tempdef}
\end{equation}
are determined by the three requirements
\begin{equation}
\left(
\begin{array}{c}
M_q\\
\bfP_q\\
\calE_q
\end{array}
\right)
=
V\Lambda^q
\int d\bfv\;
\left[
1+(q-1)\frac{m}{2k_BT}
\left|\bfv-\bfu\right|^2
\right]^{-\frac{q}{q-1}}
\left(
\begin{array}{c}
m\\
m\bfv\\
mv^2/2
\end{array}
\right),
\label{eq:reqs}
\end{equation}
and where $V\equiv\int d\bfx$ is the total spatial volume.  The
integrals can be done by making the substitution
\[
\bfw\equiv\sqrt{\frac{m}{2k_B}\left|\frac{q-1}{T}\right|}\; (\bfv-\bfu)
\]
so that
\begin{equation}
\left(
\begin{array}{c}
M_q\\
\bfP_q\\
\calE_q
\end{array}
\right)
=
mV\Lambda^q
\left(
\frac{2k_B}{m}
\left|
\frac{T}{q-1}
\right|
\right)^{D/2}
\int d\bfw\;
\left( 1+\sigma w^2\right)^{-\frac{q}{q-1}}
\left(
\begin{array}{c}
1\\
\bfu+\sqrt{\frac{2k_B}{m}\left|\frac{T}{q-1}\right|}\;\bfw\\
\frac{u^2}{2}+
\sqrt{\frac{2k_B}{m}\left|\frac{T}{q-1}\right|}\;\bfu\cdot\bfw+
\frac{k_B}{m}\left|\frac{T}{q-1}\right| w^2
\end{array}
\right),
\end{equation}
where we have defined $\sigma\equiv\sgn\left(T/\left(q-1\right)\right)
\in \{ -1,+1\}$.  The region of integration in $\bfw$ is always
spherically symmetric, so terms with odd integrands vanish, and we may
use the $D$-dimensional spherically symmetric volume element
\begin{equation}
d\bfw = \frac{2\pi^{D/2}}{\Gamma\left(\frac{D}{2}\right)}w^{D-1}dw
\label{eq:velems}
\end{equation}
on the rest, obtaining
\begin{eqnarray}
\left(
\begin{array}{c}
M_q\\
\bfP_q\\
\calE_q
\end{array}
\right)
&=&
\frac{2mV\Lambda^q}{\Gamma\left(\frac{D}{2}\right)}
\left(
\frac{2\pi k_B}{m}
\left|
\frac{T}{q-1}
\right|
\right)^{D/2}
\int dw\; w^{D-1}
\left( 1+\sigma w^2\right)^{-\frac{q}{q-1}}
\left(
\begin{array}{c}
1\\
\bfu\\
\frac{u^2}{2}+
\frac{k_B}{m}\left|\frac{T}{q-1}\right| w^2
\end{array}
\right)\nonumber\\
&=&
\frac{mV\Lambda^q}{\Gamma\left(\frac{D}{2}\right)}
\left(
\frac{2\pi k_B}{m}
\left|
\frac{T}{q-1}
\right|
\right)^{D/2}
\int dx\; x^{D/2-1}
\left( 1+\sigma x\right)^{-\frac{q}{q-1}}
\left(
\begin{array}{c}
1\\
\bfu\\
\frac{u^2}{2}+
\frac{k_B}{m}\left|\frac{T}{q-1}\right| x
\end{array}
\right),
\end{eqnarray}
where $x\equiv w^2$.

At this point, we need to specify the limits of integration.  If
$\sigma=+1$, the integrand is defined for $0\leq x < \infty$.  If
$\sigma=-1$, then the integrand is defined only for $0\leq x\leq 1$.  In
the latter case, states with $x>1$ or
\begin{equation}
\left|\bfv-\bfu\right| >
\sqrt{\frac{2k_B}{m}\left|\frac{T}{q-1}\right|}
\end{equation}
are {\it thermally forbidden}.  Thus, we have
\begin{equation}
\left(
\begin{array}{c}
M_q\\
\bfP_q\\
\calE_q
\end{array}
\right)
=
\frac{mV\Lambda^q}{\Gamma\left(\frac{D}{2}\right)}
\left(
\frac{2\pi k_B}{m}
\left|
\frac{T}{q-1}
\right|
\right)^{D/2}
\left\{
\begin{array}{ll}
\int_0^{\infty\phantom{1}} dx\; x^{D/2-1}
\left( 1+ x\right)^{-\frac{q}{q-1}}
\left(
\begin{array}{c}
1\\
\bfu\\
\frac{u^2}{2}+
\frac{k_B}{m}\left|\frac{T}{q-1}\right| x
\end{array}
\right) & \mbox{for $\sigma=+1$}\\
\int_0^{1\phantom{\infty}} dx\; x^{D/2-1}
\left( 1- x\right)^{-\frac{q}{q-1}}
\left(
\begin{array}{c}
1\\
\bfu\\
\frac{u^2}{2}+
\frac{k_B}{m}\left|\frac{T}{q-1}\right| x
\end{array}
\right) & \mbox{for $\sigma=-1$}
\end{array}
\right.
\end{equation}
In both cases, the integration results in a beta function, and this can
be expressed in terms of gamma functions~\cite{bib:as}.  The results are
\begin{equation}
\left(
\begin{array}{c}
M_q\\
\bfP_q\\
\calE_q
\end{array}
\right)
=
mV\Lambda^q
\left|
\frac{2\pi k_BT}{m}
\right|^{D/2}
\left\{
\begin{array}{ll}
\frac{\Gamma\left(\frac{q}{q-1}-\frac{D}{2}\right)}
{\left|q-1\right|^{D/2}
\Gamma\left(\frac{q}{q-1}\right)}
\left(
\begin{array}{c}
1\\
\bfu\\
\frac{u^2}{2}+\frac{Dk_BT/(2m)}{q-(q-1)\left(\frac{D}{2}+1\right)}
\end{array}
\right) & \mbox{for $\sigma=+1$}\\
\frac{\Gamma\left(\frac{1}{1-q}\right)}
{\left|1-q\right|^{D/2}
\Gamma\left(\frac{1}{1-q}+\frac{D}{2}\right)}
\left(
\begin{array}{c}
1\\
\bfu\\
\frac{u^2}{2}+\frac{Dk_BT/(2m)}{1+(1-q)\frac{D}{2}}
\end{array}
\right) & \mbox{for $\sigma=-1$}
\end{array}
\right.
\end{equation}
In either case, we see that the momentum per unit mass, or hydrodynamic
velocity, is given by
\begin{equation}
\bfu_q\equiv\frac{\bfP_q}{M_q}=\bfu.
\end{equation}
The energy per unit mass is then
\begin{equation}
\eps_q\equiv\frac{\calE_q}{M_q}=
\frac{u_q^2}{2}+\iota_q,
\end{equation}
where we have defined the thermal or {\it internal energy per unit
mass},
\begin{equation}
\iota_q\equiv
\frac{Dk_BT}{2m}
\left[ 1+\left(1-q\right)\frac{D}{2}\right]^{-1}
\label{eq:iotadef}
\end{equation}
which, remarkably, is independent of $\sigma$.

\begin{figure}
\center{\mbox{\epsfxsize=300pt\epsfbox[72 230 540 570]{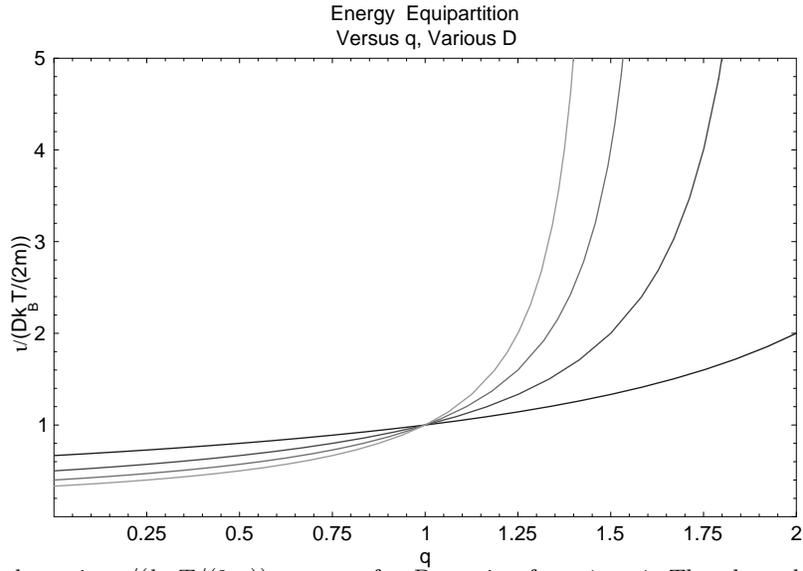}}}
\caption{A plot of the ratio $\iota_q/(k_BT/(2m))$ versus $q$ for $D$
ranging from $1$ to $4$.  The plot color becomes lighter as $D$
increases.}
\label{fig:fig1}
\end{figure}

\begin{figure}
\center{\mbox{\epsfxsize=300pt\epsfbox[72 220 540 570]{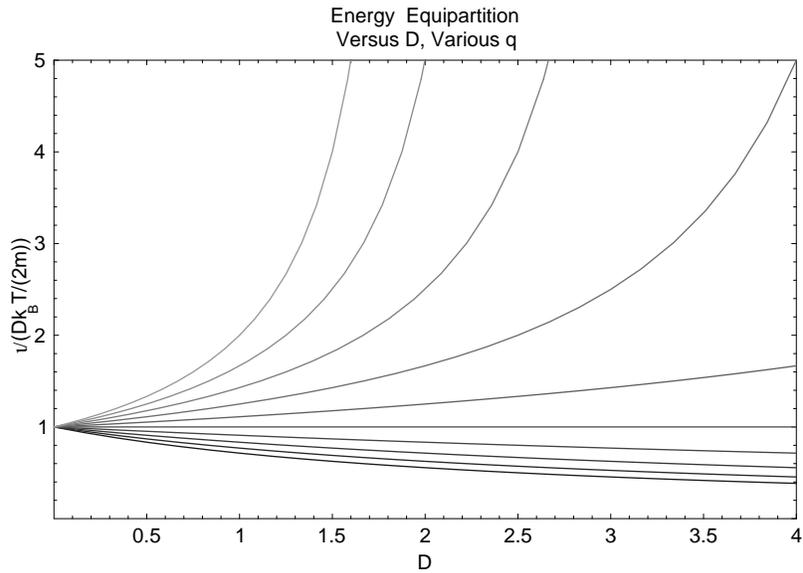}}}
\caption{A plot of the ratio $\iota_q/(k_BT/(2m))$ versus $D$ for $q$
ranging from $0.2$ to $2.0$ in increments of $0.2$.  The plot color
becomes lighter as $q$ increases.}
\label{fig:fig2}
\end{figure}

At this point, we arrive at an interesting ambiguity.  We are tempted to
interpret the variable that has been suggestively called $T$ as the {\it
temperature} of our physical system.  Certainly, it reduces to the usual
temperature for $q=1$.  If we make this interpretation, we might
conclude that the familiar energy equipartition theorem is invalid for
$q\neq 1$.  That is, for general $q$ the thermal energy per unit mass is
apparently no longer directly proportional to the dimension $D$.
Figs.~(\ref{fig:fig1}, \ref{fig:fig2}) plot the ratio
$\iota_q/(Dk_BT/(2m))$ versus $(q,D)$ for various values of $(D,q)$,
respectively.  This ratio is equal to unity when equipartition holds.
We see that it is less than unity for $q<1$, and greater than unity for
$q>1$.  We may even be tempted to attach physical significance to this
result by noting that the equipartition theorem presumes ergodicity, and
systems with $q\neq 1$ are generally not ergodic.  This reasoning is
faulty, however, because it depends upon our rather arbitrary definition
of the temperature in terms of the Lagrange multipliers in
Eq.~(\ref{eq:tempdef}).  If we had instead defined the temperature as
\begin{equation}
T'\equiv\frac{T}{1+\left(1-q\right)\frac{D}{2}},
\end{equation}
which also reduces to the usual definition as $q\rightarrow 1$, we would
have found that $\iota_q=Dk_BT'/(2m)$ and concluded that equipartition
is valid for all $q$.  The lesson here is that the definition of the
temperature is rather arbitrary in thermodynamics, and we should not
suppose that we can measure its absolute value in an experiment.  So,
even though we have theoretical reason to believe that $\iota_q(T)$ is
$q$-dependent, we can not exploit this dependence to measure $q$
experimentally.  To do that, it will be necessary to find {\it two}
physical observables whose relationship to {\it each other} is
$q$-dependent.  We shall return to this point numerous times below.

We also note from Eq.~(\ref{eq:iotadef}) that there is a hard upper
bound on $q$ for this type of system.  In order for the proportionality
constant between $\iota_q$ and $T$ not to become negative, it must be
that $1+(1-q)D/2>0$, or
\begin{equation}
q<1+\frac{2}{D}.
\label{eq:upper1}
\end{equation}
This conclusion is independent of diffeomorphic transformations in our
definition of $T$ and therefore may be expected to have some general
validity~\footnote{Of course, it could well be that we want to
generalize our notion of temperature to include negative values, in
which case this upper bound no longer applies.}.  A similar inequality
was derived in~\cite{bib:ppt}.  We shall show below that hydrodynamic
considerations -- in particular, the positivity and finiteness of the
thermal conductivity -- impose an even more stringent upper bound on
$q$.

\begin{figure}
\center{\mbox{\epsfxsize=300pt\epsfbox[72 230 540 560]{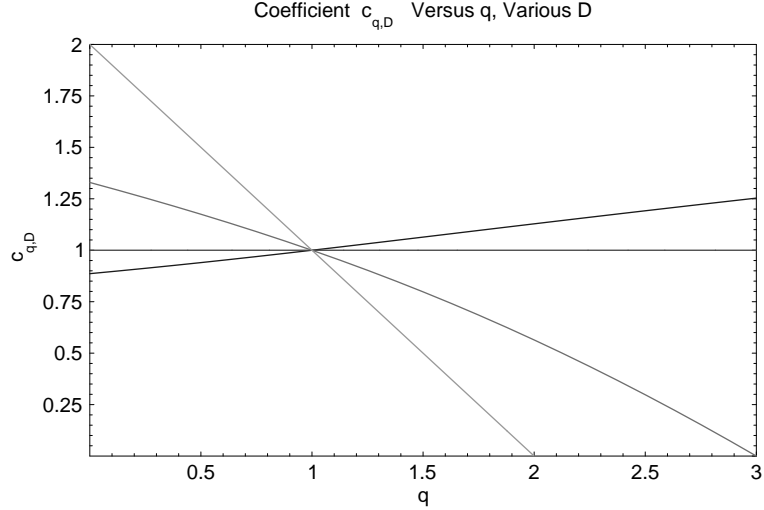}}}
\caption{A plot of the coefficients $c_{q,D}$ versus $q$ for $D$ ranging
from $1$ to $4$.  The plot color becomes lighter as $D$ increases.}
\label{fig:fig3}
\end{figure}

\begin{figure}
\center{\mbox{\epsfxsize=300pt\epsfbox[72 230 540 560]{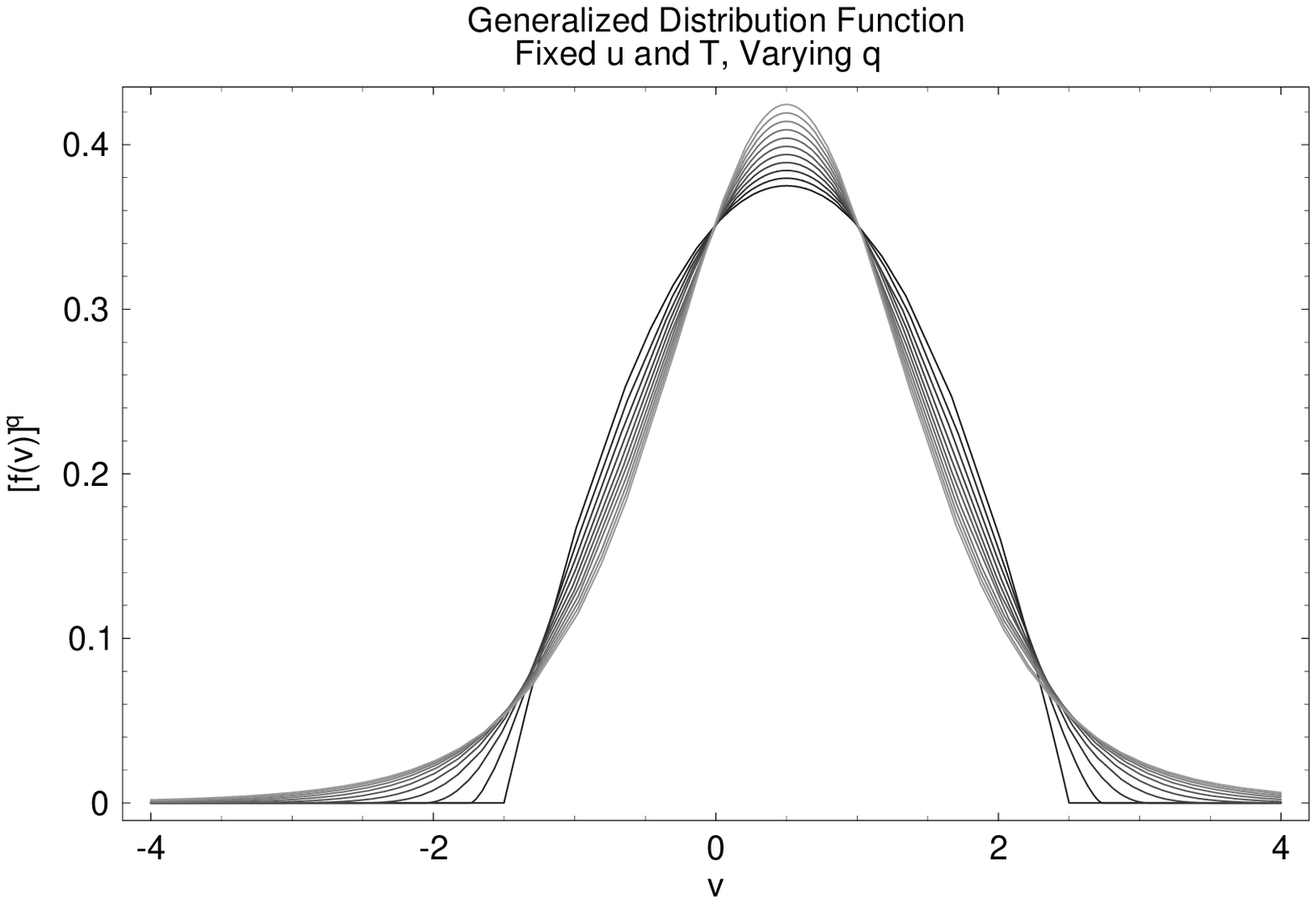}}}
\caption{A plot of $[\feq(\bfz)]^q$ versus $v$ for $D=1$, $\rho_q=1$,
$u_q=1/2$, $T=1$, and $q$ ranging from $0.5$ to $1.5$ in increments of
$0.1$.  The plot color becomes lighter as $q$ increases.}
\label{fig:fig4}
\end{figure}

\begin{figure}
\center{\mbox{\epsfxsize=300pt\epsfbox[72 230 540 560]{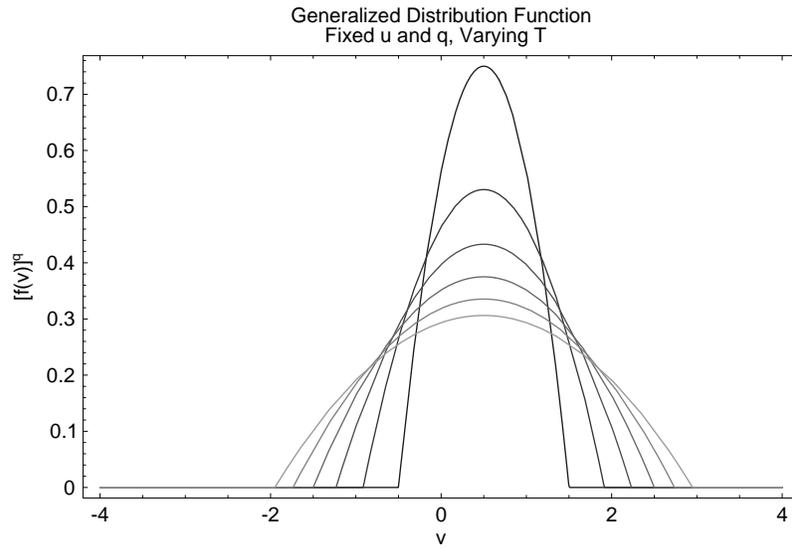}}}
\caption{A plot of the equilibrium distribution function versus $v$ for
$D=1$, $\rho_q=1$, $u_q=1/2$, $q=1/2$ and $T$ ranging from $0.25$ to
$1.5$ in increments of $0.25$.  The plot color becomes lighter as $T$
increases.}
\label{fig:fig5}
\end{figure}

Finally, the normalization constant $\Lambda$ can be expressed in terms
of the mass density $\rho_q \equiv \frac{M_q}{V}$ as follows
\begin{equation}
\Lambda^q= \frac{\rho_q}{m}
\left|\frac{m}{2\pi k_BT}\right|^{D/2}
\left\{
\begin{array}{ll}
\frac{
\left| q-1\right|^{D/2}
\Gamma\left(\frac{q}{q-1}\right)}
{\Gamma\left(\frac{q}{q-1}-\frac{D}{2}\right)} &
\mbox{for $\sigma=+1$}\\
&\\
\frac{
\left| 1-q\right|^{D/2}
\Gamma\left(\frac{1}{1-q}+\frac{D}{2}\right)}
{\Gamma\left(\frac{1}{1-q}\right)} &
\mbox{for $\sigma=-1$}
\end{array}
\right.
\end{equation}
It follows that the global hydrodynamic equilibrium (GHE) distribution
function can be written in the form
\begin{equation}
\feq (\bfz) =
\left[
\frac{c_{q,D}\rho_q}{m}
\left|\frac{m}{2\pi k_BT}\right|^{D/2}
\right]^{\frac{1}{q}}
\left[
1+\sigma\left|\frac{q-1}{k_BT}\right|
\frac{m}{2}
\left|\bfv-\bfu_q\right|^2
\right]^{-\frac{1}{q-1}},
\label{eq:eqdist}
\end{equation}
where we have defined
\begin{equation}
c_{q,D}\equiv
\left\{
\begin{array}{ll}
\frac{
\left| q-1\right|^{D/2}
\Gamma\left(\frac{q}{q-1}\right)}
{\Gamma\left(\frac{q}{q-1}-\frac{D}{2}\right)} &
\mbox{for $\sigma=+1$}\\
&\\
\frac{
\left| 1-q\right|^{D/2}
\Gamma\left(\frac{1}{1-q}+\frac{D}{2}\right)}
{\Gamma\left(\frac{1}{1-q}\right)} &
\mbox{for $\sigma=-1$,}
\end{array}
\right.
\end{equation}
and with the proviso that $\feq (\bfz)=0$ if the argument raised to the
$-1/(q-1)$ power in Eq.~(\ref{eq:eqdist}) is negative.  This is the
generalization of the Maxwell-Boltzmann distribution function for the
Generalized Thermostatistics.  The coefficients $c_{q,D}$ have a
particularly simple form for even dimension $D$,
\begin{equation}
c_{q,D}=\prod_{\ell=1}^{D/2}\left[\ell-(\ell-1)q\right];
\end{equation}
in particular, we note that $c_{q,2}=1$ and $c_{q,4}=2-q$.  More
generally, these coefficients are plotted against $q$ for various values
of $D$ in Fig.~\ref{fig:fig3}.  Various distributions for $D=1$ are
illustrated in Figs.~\ref{fig:fig4} and \ref{fig:fig5}.

\section{Generalized Kinetics}
\subsection{Local Hydrodynamic Equilibria}
\label{ssec:lhe}

Before introducing our generalization of the BGK collision operator, we
first introduce the concept of {\it local hydrodynamic equilibrium}
(LHE).  The GHE distribution function derived in the preceeding section
is a function of velocity $\bfv$ alone, and independent of position
$\bfx$.  This is a direct consequence of the fact that the only
conserved quantities present -- mass, momentum and kinetic energy -- are
functions of $\bfv$ and not of $\bfx$.  The GHE distribution does depend
parametrically on $\rho_q$, $\bfu_q$ and $\eps_q$, so we may write it in
the more precise format $\feq (\bfv;\rho_q,\bfu_q,\eps_q)$.  The LHE
distribution function $f^{(0)}(\bfz)$ is then defined as having the same
functional form as the GHE, but with parameters $\rho_q$, $\bfu_q$ and
$\eps_q$ weakly dependent on spatial position $\bfx$.  That is,
\begin{equation}
f^{(0)}
(\bfz)
=
\feq 
(\bfv;\rho_q(\bfx),\bfu_q(\bfx),\eps_q(\bfx)).
\label{eq:lhe}
\end{equation}
It should be noted that the LHE, unlike the GHE, is {\it not} expected
to be a stationary state of the system.  If the system is initialized in
a LHE distribution, its time evolution will generally take it away from
this simple functional form.  That is, the full solution to the system's
kinetic equation should be expected to be the LHE plus a correction.  To
the extent that the spatial gradients are weak, this correction will be
small and can be treated perturbatively.  This is the basis of the
Chapman-Enskog expansion.

\subsection{The Generalized BGK Equation and $H$-Theorem}

Armed with the concept of LHE, we propose the following generalization
of the Boltzmann equation with the BGK collision operator:
\begin{equation}
\left(
\frac{\partial}{\partial t} +
\bfv\cdot\bfnabla
\right) \left[f(\bfz)\right]^q =
-\frac{1}{\tau}
\left\{
\left[f(\bfz)\right]^q -
\left[f^{(0)}(\bfz)\right]^q
\right\},
\label{eq:bgk}
\end{equation}
where $f^{(0)}(\bfz)$ is the LHE distribution function that is defined
as having the same moments as the exact distribution function $f(\bfz)$,
\begin{equation}
\left(
\begin{array}{c}
\rho_q \\
\rho_q\bfu_q \\
\rho_q\eps_q
\end{array}
\right)
=
\int d\bfv\;\left[ f(\bfz) \right]^q
\left(
\begin{array}{c}
m \\
m\bfv \\
mv^2/2
\end{array}
\right)
=
\int d\bfv\;\left[ f^{(0)}(\bfz) \right]^q
\left(
\begin{array}{c}
m \\
m\bfv \\
mv^2/2
\end{array}
\right),
\label{eq:moments}
\end{equation}
and where $\tau>0$ is the collisional relaxation time.  In spite of the
deceptively simple appearance of Eq.~(\ref{eq:bgk}), note that it is
nonlinear even when $q=1$.  This is because of Eq.~(\ref{eq:moments})
which requires that the moments of $f^{(0)}(\bfz)$, which determine the
parameters that appear nonlinearly in its functional form, are
themselves functionals of $f(\bfz)$.  To see this more clearly, note
that Eqs.~(\ref{eq:lhe}) and (\ref{eq:moments}) can be used to rewrite
Eq.~(\ref{eq:bgk}) in the explicitly nonlinear but woefully cumbersome
form
\begin{equation}
\left(
\frac{\partial}{\partial t} +
\bfv\cdot\bfnabla
\right) \left[f(\bfz)\right]^q =
-\frac{1}{\tau}
\left\{
\left[f(\bfz)\right]^q -
\left[\feq\left(\bfv;\int d\bfv\; [f(\bfz)]^q m,
                \int d\bfv\; [f(\bfz)]^q m\bfv,
                \int d\bfv\; [f(\bfz)]^q \frac{m}{2}v^2\right)\right]^q
\right\}.
\end{equation}
Hence, when $q\neq 1$, this departs from the usual BGK equation in two
important respects: First, and most obviously, it is constructed so that
it relaxes to the explicitly $q$-dependent LHE $f^{(0)}(\bfz)$, given by
Eqs.~(\ref{eq:lhe}) and (\ref{eq:eqdist}), rather than to the usual
Maxwell-Boltzmann LHE.  Second, the nonlinearity in the collision
operator arises via the functional form of $[f^{(0)}]^q$ which also
explicitly depends on $q$.  It is important to keep these two mechanisms
clearly in mind because a superficial glance at Eq.~(\ref{eq:bgk}) might
suggest that the substitution $F(\bfz)\equiv [f(\bfz)]^q$ will render
its dynamics identical to those of the usual BGK equation, and this is
not so.

\begin{figure}
\center{\mbox{\epsfxsize=300pt\epsfbox[72 245 540 550]{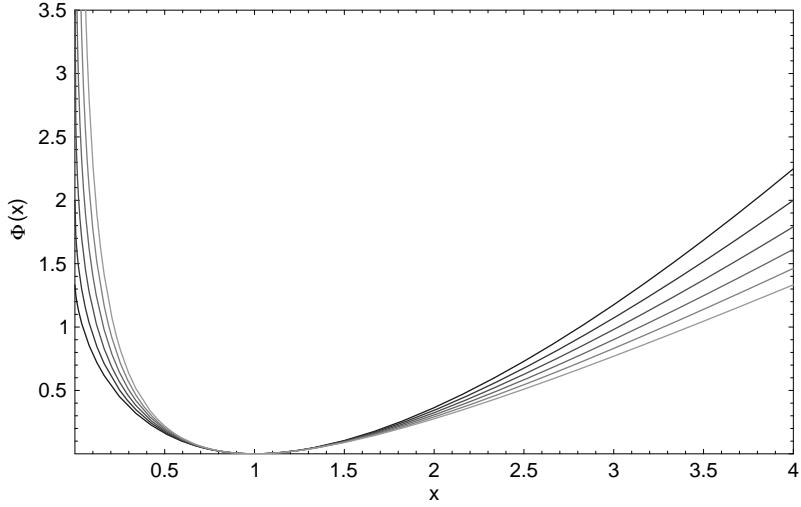}}}
\caption{A plot of the function $\Phi_q (x)$ versus $x$ for
$q\in\{0.25,0.50,0.75,1.00,1.25,1.50\}$.  The plot color becomes lighter
as $q$ increases.}
\label{fig:fig6}
\end{figure}

Eq.~(\ref{eq:bgk}) is a sensible generalization of the BGK equation for
three reasons.  First, it obviously reduces to the usual BGK equation
when $q=1$.  Second, if we multiply it by the $\gamma^i(\bfz)$ and
integrate over $\bfv$, we find local conservation equations
\begin{equation}
0 = \frac{\partial}{\partial t}
\left(
\begin{array}{c}
\rho_q(\bfx)\\
\bfp_q(\bfx)\\
\eps_q(\bfx)
\end{array}
\right) +
\bfnabla\cdot
\int d\bfv\;\left[f(z)\right]^q
\left(
\begin{array}{c}
m\bfv\\
m\bfv\bfv\\
m\bfv v^2/2
\end{array}
\right)
\end{equation}
for each of the conserved quantities.  Third, we can show that
Eq.~(\ref{eq:bgk}) obeys an $H$-theorem -- at least for $q>0$ -- as
follows:
\begin{eqnarray}
\frac{d}{dt} S_q[f]
&=&
\int d\bfz\;
\left\{\frac{\delta S_q[f]}{\delta f(z)}\right\}
\frac{\partial f(\bfz)}{\partial t}\nonumber\\
&=&
\frac{k_B}{q-1}
\int d\bfz\;
\left\{1-q[f(\bfz)]^{q-1}\right\}
\frac{\partial f(\bfz)}{\partial t}\nonumber\\
&=&
\frac{k_B}{q-1}
\int d\bfz\;
\left\{\frac{1}{q[f(\bfz)]^{q-1}}-1\right\}
\frac{\partial}{\partial t}
[f(\bfz)]^q\nonumber\\
&=&
\frac{k_B}{q-1}
\int d\bfz\;
\left\{\frac{1}{q[f(\bfz)]^{q-1}}-1\right\}
\left(\frac{\partial}{\partial t}
+\bfv\cdot\bfnabla\right)
[f(\bfz)]^q\nonumber\\
&=&
-\frac{k_B}{(q-1)\tau}
\int d\bfz\;
\left\{\frac{1}{q[f(\bfz)]^{q-1}}-1\right\}
\left\{[f(\bfz)]^q - [f^{(0)}(\bfz)]^q\right\}\nonumber\\
&=&
\frac{S_q\left[f^{(0)}\right] - S_q\left[f\right]}{\tau}
+\frac{k_B}{(q-1)\tau}
\int d\bfz\;
\left\{\left[f(\bfz)-f^{(0)}(\bfz)\right]-
\frac{[f(\bfz)]^q-[f^{(0)}(\bfz)]^q}{q[f(\bfz)]^{q-1}}\right\}\nonumber\\
&=&
\frac{S_q\left[f^{(0)}\right] - S_q\left[f\right]}{\tau}
+\frac{k_B}{(q-1)\tau}
\int d\bfz\; f^{(0)}(\bfz)
\left\{\left[g(\bfz)-1\right]-
\frac{[g(\bfz)]^q-1}{q[g(\bfz)]^{q-1}}\right\}\nonumber\\
&=&
\frac{S_q\left[f^{(0)}\right] - S_q\left[f\right]}{\tau}
+\frac{k_B}{\tau}
\int d\bfz\; f^{(0)}(\bfz)
\Phi_q\left(g(\bfz)\right),
\label{eq:hth}
\end{eqnarray}
where $g(\bfz)\equiv f(\bfz)/f^{(0)}(\bfz) > 0$, the domain has been
assumed to be without boundary, and where we have defined the function
\begin{equation}
\Phi_q(x)\equiv \frac{1}{q-1}
\left[
\left(x-1\right)-\frac{x^q-1}{qx^{q-1}}
\right].
\end{equation}
The first term of Eq.~(\ref{eq:hth}) is nonnegative since the entropy is
a maximum for the distribution function $f^{(0)}(\bfz)$ and $q>0$ by
construction.  It is not difficult to show that the function $\Phi_q$ is
nonnegative for positive argument (see Fig.~\ref{fig:fig6}), so the
second term is also nonnegative.  It follows that $dS_q[f]/dt \geq 0$
for $q>0$, with equality holding only at equilibrium.

For $q<0$, the generalized entropy $S_q[f]$ is {\it minimized} at its
extremum, rather than maximized, as can be seen from examination of its
second variation.  In this situation, the first term on the right-hand
side of Eq.~(\ref{eq:hth}) is negative, but the second term is still
positive because $\Phi_q(x)\geq 0$ still holds for $q\leq 0$.  Hence the
above argument breaks down and a more subtle one is needed.  We shall
not concern ourselves with the case $q\leq 0$ in this paper.

\section{The Chapman-Enskog Analysis}
\subsection{The Asymptotic Expansion}

We now develop the distribution function in a perturbation series where
the zero-order approximation is the LHE, following the program described
in Subsection~\ref{ssec:lhe}.  We rewrite our generalized BGK equation,
Eq.~(\ref{eq:bgk}), in the form
\begin{equation}
\left(
1+\tau\calD
\right) F(\bfz) = \Fnt (\bfz),
\end{equation}
where we have defined
\begin{eqnarray}
F(\bfz) &\equiv& \left[f(\bfz)\right]^q\\
\noalign{\noindent and}\\
\Fnt(\bfz) &\equiv& \left[\fnt(\bfz)\right]^q,
\end{eqnarray}
and the operator
\begin{equation}
\calD\equiv
\frac{\partial}{\partial t}+
\bfv\cdot\bfnabla.
\end{equation}
The formal solution to the above equation is
\begin{equation}
F(\bfz) =
\left(
1+\tau\calD
\right)^{-1} \Fnt (\bfz).
\end{equation}
Following the standard Chapman-Enskog procedure~\cite{bib:ce}, we expand
the derivative in a series of successively more slowly varying time
scales,
\begin{equation}
\calD = \sum_{j=1}^\infty \epsilon^j\calD_j,
\end{equation}
where
\begin{equation}
\calD_j\equiv
\frac{\partial}{\partial t_j}+
\bfv\cdot\bfnabla,
\end{equation}
where $\epsilon$ is a formal expansion parameter, and $t_j$ is the $j$th
time scale.  To second order in $\epsilon$, we find
\begin{equation}
F(\bfz) =
\left[
1-\epsilon\tau\calD_1-\epsilon^2\tau
\left(\calD_2-\tau\calD_1^2\right)+
\cdots\right]\Fnt(\bfz).
\end{equation}
Thus we have
\begin{equation}
F(\bfz) = \sum_{n=0}^\infty\epsilon^n \Fnn{n}(\bfz),
\end{equation}
where
\begin{eqnarray}
\Fnn{1}(\bfz) &=& -\tau\calD_1\Fnt(\bfz)
\label{eq:frsto}\\
\Fnn{2}(\bfz) &=& -\tau\left(\calD_2-\tau\calD_1^2\right)\Fnt(\bfz).
\label{eq:scndo}\\
&\vdots& \nonumber
\end{eqnarray}
The higher order terms $\Fnn{1}$ and $\Fnn{2}$ must not change the
definition of the conserved densities,
\begin{equation}
0 =
\int d\bfv\;\left[F(\bfz)-\Fnt(\bfz)\right]
\left(
\begin{array}{c}
m\bfv\\
m\bfv\bfv\\
m\bfv v^2/2
\end{array}
\right)
= \sum_{n=1}^\infty\epsilon^n
\int d\bfv\;\Fnn{n}(\bfz)
\left(
\begin{array}{c}
m\bfv\\
m\bfv\bfv\\
m\bfv v^2/2
\end{array}
\right).
\label{eq:ccinv}
\end{equation}
To fully specify the determination of the $\Fnn{n}$, we further require
that they satisfy Eq.~(\ref{eq:ccinv}) order by order; that is,
\begin{equation}
0 =
\int d\bfv\;\Fnn{n}(\bfz)
\left(
\begin{array}{c}
m\bfv\\
m\bfv\bfv\\
m\bfv v^2/2
\end{array}
\right)
\label{eq:ccinvoo}
\end{equation}
for $n\geq 1$.

\subsection{The First-Order Solution}

At first order, Eq.~(\ref{eq:frsto}) can be written
\begin{equation}
\Fnn{1}(\bfz) = -\tau
\left(
\frac{\partial}{\partial t_1}+\bfv\cdot\bfnabla
\right)
\Fnt(\bfz).
\end{equation}
We take the hydrodynamic moments of both sides, noting that the
left-hand side will vanish thanks to Eq.~(\ref{eq:ccinvoo}).  We get
\begin{equation}
0 =
\frac{\partial}{\partial t_1}
\left(
\begin{array}{c}
\rho_q\\
\rho_q\bfu_q\\
\rho_q\eps_q
\end{array}
\right) +
\bfnabla\cdot
\int d\bfv\;
\Fnt(\bfz)
\left(
\begin{array}{c}
m\bfv\\
m\bfv\bfv\\
m\bfv v^2/2
\end{array}
\right).
\end{equation}

The first of these is immediately seen to be the usual equation
expressing conservation of mass
\begin{equation}
0 = \frac{\partial\rho_q}{\partial t_1} +
\bfnabla\cdot\left(\rho_q\bfu_q\right),
\label{eq:mass1}
\end{equation}
which is thus seen to be $q$-invariant at first order.

To evaluate the momentum equation at first order, we must integrate the
dyad $m\bfv\bfv$ times $\Fnt$.  We have
\begin{eqnarray}
\int d\bfv\; \Fnt(\bfz) m\bfv\bfv
&=&
\int d\bfv\; \Fnt(\bfz) m
\left[\left(\bfv-\bfu_q\right)+\bfu_q\right]
\left[\left(\bfv-\bfu_q\right)+\bfu_q\right]\\
&=&
\rho_q\bfu_q\bfu_q +
\left(\frac{2k_B}{m}\left|\frac{T}{q-1}\right|\right)^{\frac{D}{2}+1}
\int d\bfw\; \Fnt(\bfz) m\bfw\bfw,
\end{eqnarray}
since the odd cross terms integrate to zero.  Turning our attention to
the last integral over $\bfw$, it is clear from parity that only the
diagonal elements of this dyad will have nonvanishing integral, and from
isotropy that they will all give the same result.  Thus, we have
\begin{equation}
\int d\bfv\; \Fnt(\bfz) m\bfv\bfv =
\rho_q\bfu_q\bfu_q +
\bfone\left(\frac{2k_B}{m}\left|\frac{T}{q-1}\right|\right)^{\frac{D}{2}+1}
\int d\bfw\; \Fnt(\bfz) mw_x^2.
\end{equation}
To perform this last integral, the spherically symmetric volume element
of Eq.~(\ref{eq:velems}) is inadequate.  Rather, we write
$w_x=w\cos\theta$, where $0\leq\theta\leq\pi$ is a polar angle, and we
adopt the cylindrically symmetric volume element
\begin{equation}
d\bfw =
\frac{2\pi^{\frac{D-1}{2}}}{\Gamma\left(\frac{D-1}{2}\right)}
w^{D-1} dw\; d\theta.
\end{equation}
Now both the $w$ and $\theta$ integrals yield beta-functions which must
be considered for both cases $\sigma=\pm 1$.  The result, which does not
depend on $\sigma$, is then
\begin{equation}
\int d\bfv\; \Fnt(\bfz) m\bfv\bfv =
\rho_q\bfu_q\bfu_q +
P_q \bfone
\end{equation}
where we have defined the {\it pressure}
\begin{equation}
P_q\equiv\frac{\rho_q k_BT}{m\left[1+(1-q)\frac{D}{2}\right]}.
\label{eq:pdef}
\end{equation}
The momentum equation at first order is thus
\begin{equation}
0 =
\frac{\partial}{\partial t_1}
\left(\rho_q\bfu_q\right)+
\bfnabla\cdot
\left(
\rho_q\bfu_q\bfu_q + P_q\bfone
\right).
\label{eq:mom1}
\end{equation}

Eq.~(\ref{eq:pdef}) appears to be a nonideal equation of state for the
pressure but, as noted at the end of the previous subsection, we must
avoid attaching physical significance to the temperature $T$.  Rather,
we should take care to relate observable quantities -- in this case, the
pressure $P_q$ and the internal energy $\iota_q$.  From
Eqs.~(\ref{eq:pdef}) and (\ref{eq:iotadef}), we have
\begin{equation}
P_q = \frac{2}{D}\rho_q\iota_q,
\label{eq:eos}
\end{equation}
and this is the usual ideal gas equation of state with no correction,
which we see is $q$-invariant after all.

Finally, we consider the energy equation at first order.  We must
integrate the vector $m\bfv v^2/2$ times $\Fnt$.  We have
\begin{eqnarray}
\int d\bfv\; \Fnt(\bfz) \frac{m\bfv v^2}{2}
&=&
\int d\bfv\; \Fnt(\bfz) \frac{m}{2}
\left[\left(\bfv-\bfu_q\right)+\bfu_q\right]
\left|\left(\bfv-\bfu_q\right)+\bfu_q\right|^2\\
&=&
\rho_q\eps_q\bfu_q + \bfu_q\cdot
\left[
\left(\frac{2k_B}{m}\left|\frac{T}{q-1}\right|\right)^{\frac{D}{2}+1}
\int d\bfw\; \Fnt(\bfz) m\bfw\bfw
\right]
\end{eqnarray}
since the odd cross terms integrate to zero.  The quantity in square
brackets in the second term is the same one we encountered in the
definition of the pressure and is equal to $P_q\bfone$.  The energy
equation at first order is thus
\begin{equation}
0 =
\frac{\partial}{\partial t_1}
\left(\rho_q\eps_q\right)+
\bfnabla\cdot
\left[
\left(
\rho_q\eps_q + P_q
\right)\bfu_q
\right].
\label{eq:energy1}
\end{equation}

Eqs.~(\ref{eq:mass1}), (\ref{eq:mom1}) and (\ref{eq:energy1}) give us
the rate of change of the hydrodynamic variables on time scale $t_1$.
They and the equation of state, Eq.~(\ref{eq:eos}), are all seen to be
$q$-invariant.  Thus, at first order, there is no experiment that we
could perform that would distinguish a system with $q\neq 1$ from the
more orthodox $q=1$ case.  To make this distinction, we must examine the
dissipative terms that appear at second order in the Chapman-Enskog
analysis, and we turn our attention to those now.

\subsection{The Second-Order Solution}

At second order, Eq.~(\ref{eq:scndo}) can be written
\begin{equation}
\Fnn{2}(\bfz) = -\tau\left[
\left(
\frac{\partial}{\partial t_2}+\bfv\cdot\bfnabla
\right)\Fnt(\bfz)
-\tau
\left(
\frac{\partial}{\partial t_1}+\bfv\cdot\bfnabla
\right)^2\Fnt(\bfz)
\right].
\end{equation}
Again, we take the hydrodynamic moments of both sides, noting that the
left-hand side will vanish thanks to Eq.~(\ref{eq:ccinvoo}).  Drawing on
our experience from the first-order calculation, we see that we get
\begin{eqnarray}
\lefteqn{
\frac{\partial}{\partial t_2}
\left(
\begin{array}{c}
\rho_q\\
\rho_q\bfu_q\\
\rho_q\eps_q
\end{array}
\right) +
\bfnabla\cdot
\left(
\begin{array}{c}
\rho_q\bfu_q\\
\rho_q\bfu_q\bfu_q+P_q\bfone\\
\left(
\rho_q\eps_q + P_q
\right)\bfu_q
\end{array}
\right)}\nonumber\\
&=&
\tau
\left[
\frac{\partial^2}{\partial t_1^2}
\left(
\begin{array}{c}
\rho_q\\
\rho_q\bfu_q\\
\rho_q\eps_q
\end{array}
\right) +
2\frac{\partial}{\partial t_1}\bfnabla\cdot
\left(
\begin{array}{c}
\rho_q\bfu_q\\
\rho_q\bfu_q\bfu_q+P_q\bfone\\
\left(
\rho_q\eps_q + P_q
\right)\bfu_q
\end{array}
\right) +
\bfnabla\bfnabla :
\int d\bfv\;\bfv\bfv\Fnt(\bfz)
\left(
\begin{array}{c}
m\\
m\bfv\\
m v^2/2
\end{array}
\right)
\right]\nonumber\\
&=&
\tau\bfnabla\cdot
\left[
\frac{\partial}{\partial t_1}
\left(
\begin{array}{c}
\rho_q\bfu_q\\
\rho_q\bfu_q\bfu_q+P_q\bfone\\
\left(
\rho_q\eps_q + P_q
\right)\bfu_q
\end{array}
\right) +
\bfnabla\cdot
\int d\bfv\;\bfv\bfv\Fnt(\bfz)
\left(
\begin{array}{c}
m\\
m\bfv\\
m v^2/2
\end{array}
\right)
\right],
\end{eqnarray}
where we used the first order results in the second step.  Note that the
right-hand side is now manifestly a divergence.  The quantity inside
this divergence is the negative of a diffusive flux.  The derivatives
with respect to $t_1$ can be eliminated using the first-order evolution
equations.

The right-hand side of the second-order mass conservation equation is
seen to vanish due to the first-order momentum conservation equation.
Thus, we find that the form of the mass conservation equation
\begin{equation}
0 = \frac{\partial\rho_q}{\partial t_2} +
\bfnabla\cdot\left(\rho_q\bfu_q\right)
\label{eq:mass2}
\end{equation}
is $q$-invariant to second order.

To evaluate the right-hand side of the momentum equation at second
order, we must integrate the triad $m\bfv\bfv\bfv$ times $\Fnt$.  The
procedure is the same as that at first order, and no new moments need to
be defined.  The result in index notation is
\begin{equation}
\int d\bfv\; \Fnt(\bfz) mv_iv_jv_k
=
\rho_q {u_q}_i {u_q}_j {u_q}_k +
P_q
\left(
\delta_{ij}{u_q}_k +
\delta_{ik}{u_q}_j +
\delta_{jk}{u_q}_i
\right).
\end{equation}
Inserting this into the second-order momentum equation, carrying out the
derivatives with respect to $t_1$ using the first-order equations, and
simplifying, we get
\begin{equation}
\frac{\partial}{\partial t_2}
\left(\rho_q\bfu_q\right) +
\bfnabla\cdot\left(\rho_q\bfu_q\bfu_q+P_q\bfone\right) =
\bfnabla\cdot
\left\{
\mu_q
\left[
\bfnabla\bfu_q + \left(\bfnabla\bfu_q\right)^T
\right] +
\lambda_q
\left(\bfnabla\cdot\bfu_q\right)\bfone
\right\},
\label{eq:mom2}
\end{equation}
where the superscript $T$ denotes ``transpose,'' and where we have
defined the {\it shear viscosity}
\begin{equation}
\mu_q\equiv\frac{2\tau}{D}\rho_q\iota_q,
\label{eq:viss}
\end{equation}
and the {\it bulk viscosity}
\begin{equation}
\lambda_q\equiv -\frac{4\tau}{D^2}\rho_q\iota_q.
\label{eq:visb}
\end{equation}
Here we have used the equation of state, Eq.~(\ref{eq:eos}), to express
the viscosities in terms of the internal energy.  Note that
Eqs.~(\ref{eq:viss}) and (\ref{eq:visb}) are $q$-invariant, as is the
ratio
\begin{equation}
\frac{\lambda_q}{\mu_q} = -\frac{2}{D}
\end{equation}
which is known to be approximately true for a variety of real
gases~\cite{bib:harlow}.  Thus, measurement of the ratio of the
viscosities is still insufficient to determine $q$ experimentally.

Finally, we turn our attention to the energy equation at second order.
For this we must integrate the diad $m\bfv\bfv v^2/2$ times $\Fnt$.  The
by now familiar procedure again yields beta-function integrals which
must be considered for both cases $\sigma=\pm 1$.  The result, which
does not depend on $\sigma$, is then
\begin{equation}
\int d\bfv\; \Fnt(\bfz) m\bfv\bfv\frac{v^2}{2}
=
\left(
\rho_q\eps_q + 2P_q
\right)\bfu_q\bfu_q +
\left\{
P_q\frac{u_q^2}{2} + \frac{4}{D^2}\left(\frac{D}{2}+1\right)
\left[\frac{1+(1-q)\frac{D}{2}}{1+(1-q)\left(\frac{D}{2}+1\right)}\right]
\rho_q\iota_q^2
\right\}\bfone.
\end{equation}
Inserting this into the second-order energy equation, carrying out the
derivatives with respect to $t_1$ using the first-order equations, and
simplifying, we get
\begin{eqnarray}
\lefteqn{
\frac{\partial}{\partial t_2}
\left(\rho_q\eps_q\right) +
\bfnabla\cdot\left[\left(\rho_q\eps_q+P_q\right)\bfu_q\right]}\nonumber\\
&=&
\bfnabla\cdot
\left\{
\bfu_q\cdot
\left[
\mu_q
\left(
\bfnabla\bfu_q + \left(\bfnabla\bfu_q\right)^T
\right) +
\lambda_q
\left(\bfnabla\cdot\bfu_q\right)\bfone
\right] +
k_q\bfnabla\iota_q -
(1-q)\frac{a_q}{\rho_q}\bfnabla\left(\rho_q\iota_q\right)
\right\}
\label{eq:energy2}
\end{eqnarray}
where we have defined the {\it thermal conductivity}
\begin{equation}
k_q\equiv \frac{4}{D^2}\left(1+\frac{D}{2}\right)
\left[\frac{1+(1-q)\frac{D}{2}}{1+(1-q)\left(\frac{D}{2}+1\right)}\right]
\tau\rho_q\iota_q,
\end{equation}
and the anomalous transport coefficient
\begin{equation}
a_q\equiv \frac{4}{D^2}
\left[\frac{\frac{D}{2}+1}{1+(1-q)\left(\frac{D}{2}+1\right)}\right]
\tau\rho_q\iota_q.
\end{equation}
We note that positivity and finiteness of the thermal conductivity sets
an even more stringent upper bound on $q$ than Eq.~(\ref{eq:upper1}),
namely
\begin{equation}
q < 1 + \frac{2}{D+2}.
\label{eq:upper2}
\end{equation}
We emphasize that this inequality is not expected to hold for general
systems; its derivation was specific to our assumption of an ideal gas.

\begin{figure}
\center{\mbox{\epsfxsize=300pt\epsfbox[72 230 540 560]{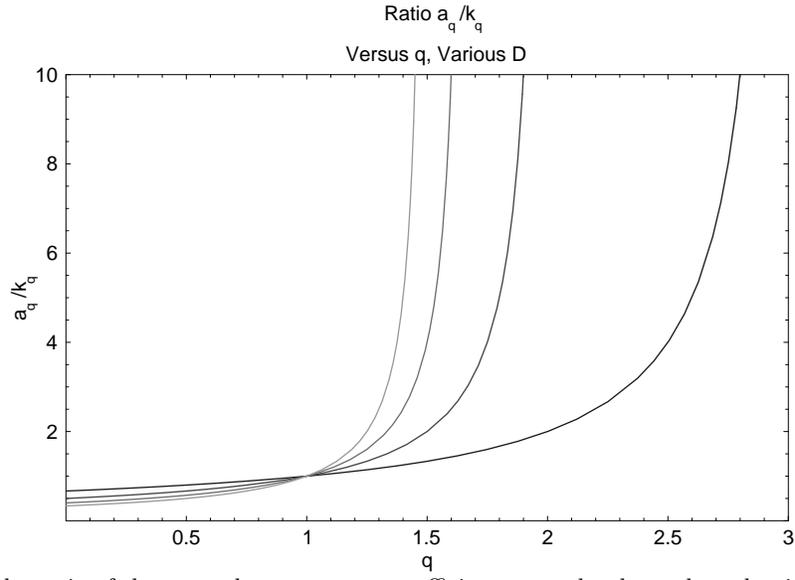}}}
\caption{A plot of the ratio of the anomalous transport coefficient
$a_q$ to the thermal conductivity $k_q$ for $D$ ranging from $1$ to $4$.
The plot color becomes lighter as $D$ increases.}
\label{fig:fig7}
\end{figure}

\begin{figure}
\center{\mbox{\epsfxsize=300pt\epsfbox[72 230 540 560]{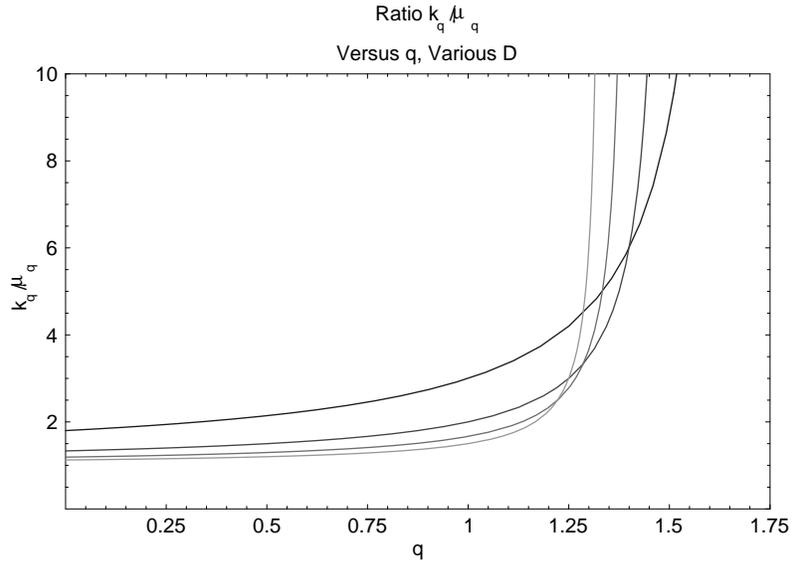}}}
\caption{A plot of the ratio of the thermal conductivity $k_q$ to the
shear viscosity $\mu_q$ for $D$ ranging from $1$ to $4$.  The plot color
becomes lighter as $D$ increases.}
\label{fig:fig8}
\end{figure}

The most striking feature of the second-order energy conservation
equation, Eq.~(\ref{eq:energy2}), is that it is {\it not} $q$-invariant.
Its diffusive flux contains a term proportional to
$(1-q)\bfnabla\left(\rho_q\iota_q\right)/\rho_q$ with a new transport
coefficient $a_q$.  The presence of this term can be detected by purely
hydrodynamic experiments and this may be used to test whether or not $q$
is equal to unity.  Moreover, we see that the ratio
\begin{equation}
\frac{a_q}{k_q} = \left[1+(1-q)\frac{D}{2}\right]^{-1}
\end{equation}
gives us another means of determining $q$, as does the ratio
\begin{equation}
\frac{k_q}{\mu_q} = \frac{2}{D}\left(\frac{D}{2}+1\right)
\left[\frac{1+(1-q)\frac{D}{2}}{1+(1-q)\left(\frac{D}{2}+1\right)}\right].
\end{equation}
These ratios are plotted in Figs.~\ref{fig:fig7} and \ref{fig:fig8}.

\section{Discussion}

Eqs.~(\ref{eq:mass2}), (\ref{eq:mom2}) and (\ref{eq:energy2}) constitute
the compressible Navier-Stokes equations for the generalized
thermostatistics.  We see that the mass and momentum conservation
equations are completely $q$-invariant, but the energy conservation
equation is not.  It contains a term that is proportional to
$(1-q)\bfnabla\left(\rho_q\iota_q\right)/\rho_q$ which vanishes for
Boltzmann-Gibbs statistics.  It is a term not normally included in
presentations of the Navier-Stokes equations.

In addition, we have shown that certain ratios of the transport
coefficients may be $q$ dependent.  While the famous result
$\lambda_q/\mu_q=-2/D$ appears to be $q$-invariant, the ratios
$k_q/\mu_q$ and $a_q/k_q$ are decidedly less robust and may be used to
infer a value of $q$.  At this point, the reader may be tempted to pick
up a handbook of material properties, search for materials with
anomalous ratios of thermal conductivity to shear viscosity, and
attribute those anomalies to a breakdown in Boltzmann-Gibbs statistics
for those materials.  {\it The reader is implored to resist this
temptation.}  There are very many potential pitfalls in the
Boltzmann/Chapman-Enskog analysis that are far more likely to generate
such anomalies than a breakdown in the foundations of thermostatistics.
The presence of rotational and/or internal molecular degrees of freedom,
high densities leading to three-body collisions, and violations of the
Boltzmann molecular chaos assumption are but three examples of phenomena
that might also cause such anomalies.  If, however, one has other
reasons to believe that a particular system violates Boltzmann-Gibbs
thermostatistics, then examination of the ratios of the transport
coefficients may provide further corroboration.  One does have such
reason to believe this for the stellar polytrope and pure-electron
plasma examples mentioned earlier; unfortunately, those systems of
particles are not ideal gases in any reasonable sense of the term so,
even if one could measure the transport coefficients of such things, the
quantitative results derived above should not be expected to apply.

In summary, the presence of the anomalous term in the energy equation of
a dilute gas would seem to be the easiest way to search for deviations
of $q$ from unity.  I am currently unaware of any experimental work that
would clearly establish either the presence or absence of this term.  I
leave it to my colleagues who have more familiarity with the
experimental hydrodynamic literature to sort out this matter.

\section{Conclusions}

We have shown that the Navier-Stokes equations for mass and momentum
conservation are $q$-invariant to second order in the Chapman-Enskog
expansion, but that the equation for energy conservation is not.  We
have derived the form of this anomalous term, using a generalized
Chapman-Enskog analysis on a generalized BGK kinetic equation.  In
addition, we have found $q$-dependent anomalies in ratios of certain
transport coefficients -- in particular, in the ratio of the thermal
conductivity to the shear viscosity.  Finally, we have found that
hydrodynamics gives rise to the upper bound $q < 1+2/(D+2)$ which is
more stringent than that imposed by equilibrium considerations.

This work could be substantially improved by using a more realistic
kinetic equation, but we have argued that the principal results -- the
{\it form} of the hydrodynamic equations, and the {\it ratios} of the
transport coefficients -- ought to be robust in this regard.  It is
hoped that some of the insights obtained in this analysis will
eventually be useful in constructing simple experimental tests for the
presence of breakdowns in Boltzmann-Gibbs statistics.

\section*{Acknowledgements}

The author would like to thank Constantino Tsallis and Lisa Borland for
their careful reading of the original draft of this paper, and for their
helpful suggestions.  The author would also like to thank Celia
Anteneodo for sharing the results of her recent recalculation of the
radial profile of the pure-electron plasma using normalized
$q$-expectation values~\cite{bib:antp}.


\begin{thebibliography}{999}
  \parindent=.6em 
\bibitem{bib:tsallis} C. Tsallis, {\it J. Stat. Phys.} {\bf 52} (1988)
 479; {\it Physica A} {\bf 221} (1995) 277.
\bibitem{bib:leg} E.M.F. Curado, C. Tsallis, {\it J. Phys. A:
 Math. Gen.} {\bf 24} L69.
\bibitem{bib:or} A. Chame, E.V.L. de Mello, {\it Phys. Lett. A} {\bf
 228} (1997) 159.
\bibitem{bib:fd} A. Chame, E.V.L. de Mello, {\it J. Phys. A} {\bf 27}
 (1994) 3663.
\bibitem{bib:pp} A.R. Plastino, A. Plastino, {\it Phys. Lett. A} {\bf
 174} (1993) 384-386.
\bibitem{bib:bog} B.M. Boghosian, {\it Phys. Rev. E} {\bf 53} (1996)
 4754-4763.
\bibitem{bib:ant} C. Anteneodo, C. Tsallis, {\it J. Mol. Liq.} {\bf 71}
 (1997) 255-267.
\bibitem{bib:levya} P.A. Alemany, D.H. Zanette, {\it Phys. Rev. E} {\bf
 49} (1994) R956.
\bibitem{bib:levyb} C. Tsallis, S.V.F. Levy, A.M.C. de Souza,
 R. Maynard, {\it Phys. Rev. Lett.} {\bf 75} (1995) 3589; Erratum: {\it
 Phys. Rev. Lett.} {\bf 77} (1996) 5442.
\bibitem{bib:levyc} C. Tsallis, {\it Physics World} (July, 1997) 42.
\bibitem{bib:ions} C. Tsallis, A.M.C. de Souza, {\it Phys. Lett. A} {\bf
 235} (1997) 444-446.
\bibitem{bib:gold} N. Goldenfeld, ``Lectures on Phase Transitions and
 the Renormalization Group,'' Addison-Wesley (1992) 27-28.
\bibitem{bib:as} M. Abramowitz and I.A. Stegun, ``Handbook of
 Mathematical Functions,'' National Bureau of Standards (ninth printing,
 1970) p. 258.
\bibitem{bib:ppt} A.R. Plastino, A. Plastino, C. Tsallis, {\it
 J. Phys. A: Math. Gen.} {\bf 27} (1994) 5707-5714.
\bibitem{bib:ce} K. Huang, ``Statistical Physics,'' John Wiley and Sons
 (third printing, 1966).
\bibitem{bib:cant} C. Tsallis, R.S. Mendes, A.R. Plastino, {\it Physica
 A} {\bf 261} (1998) 534.
\bibitem{bib:abe} S. Abe, ``Thermodynamic Limit and Classical Ideal Gas
 in Nonextensive Statistical Mechanics with Normalised $q$-Expectation
 Values,'' preprint (1998).
\bibitem{bib:antp} C. Anteneodo, private communication (1998).
\bibitem{bib:harlow} F.H. Harlow, A.A. Amsden, ``Fluid Dynamics -- A
 LASL Monograph,'' Los Alamos publication LA-4700 (1971).
\end{thebibliography}
\end{document}